\documentclass{appolb}
\usepackage{graphicx}

\usepackage[english]{babel}
\usepackage[latin1]{inputenc}

\usepackage{braket}
\usepackage{subfigure}
\renewcommand{\d}{\ensuremath{\mathrm{d}}}

\begin{document}
\title{Aspects of gluon propagation in Landau gauge: \\
spectral densities, and mass scales at finite temperature%
\thanks{Presented at the ``Workshop on Unquenched Hadron Spectroscopy: \\
     Non-Perturbative Models and Methods of QCD vs. Experiment'',
               at the occasion of Eef van Beveren's 70th birthday, 
        1-5 September 2014, University of Coimbra, Portugal}%
}
\author{Paulo J. Silva, Orlando Oliveira 
\address{Centro de F\'{i}sica Computacional, Universidade de Coimbra, Portugal} \\ 
David Dudal 
\address{KU Leuven Campus Kortrijk - KULAK, Department of Physics, Etienne Sabbelaan 53, 8500 Kortrijk, Belgium} 
\address{Department of Physics and Astronomy, Ghent University, Belgium} \\
Pedro Bicudo \address{CFTP, Instituto Superior T\'{e}cnico, Portugal} \\
Nuno Cardoso \address{NCSA, University of Illinois, Urbana IL 61801, USA} }
\maketitle
\begin{abstract}
We discuss a method to extract the K\"{a}llén-Lehmann 
spectral density of a particle (be it elementary or bound state) propagator 
and apply it to compute gluon spectral densities from lattice data. 
Furthermore, we also consider the interpretation of the Landau-gauge 
gluon propagator at finite temperature as a massive-type bosonic
propagator.
\end{abstract}
  
\section{Spectral densities}

In general, an Euclidean momentum-space propagator 
$\mathcal{G}(p^2)\equiv\braket{\mathcal{O}(p)\mathcal{O}(-p)}$ 
of a (scalar) physical degree of freedom ought to have a 
K\"{a}ll\'{e}n-Lehmann spectral representation

\begin{equation}
\mathcal{G}(p^2)=\int_{0}^{\infty}\d\mu\frac{\rho(\mu)}{p^2+\mu}
\,.
\label{specdens}
\end{equation}

The knowledge of the spectral function $\rho(\mu)$ is useful for, amongst 
other things, to get the masses of the physical states described by the operator
$\mathcal{O}$.

Here we describe a method to compute the spectral density given a numerical 
estimate of the propagator, computed using e.g. lattice techniques.
Note that eq. (\ref{specdens}) is equivalent to a double Laplace transform
$  \mathcal{G}=\mathcal{L}^2\hat\rho=\mathcal{L}\mathcal{L}^\ast \hat\rho$
where $(\mathcal{L}f)(t)\equiv\int_0^{\infty}\d s e^{-st}f(s)$; the (double) inversion is then a notorious ill-posed problem, due to the exponential dampening.

For positive spectral functions, a popular approach is the maximum entropy 
method \cite{mem}. An alternative approach, aiming to compute spectral 
densities not necessarily positive, has been developed by 
some of us \cite{letter}, based on the Tikhonov regularization supplemented with the Morozov discrepancy principle. 

Specifically,
setting $D_i\equiv D(p_i^2)$ and assuming we have $N$ data points, we minimize
\begin{equation}\label{tikdis1}
  \mathcal{J}_\lambda=\sum_{i=1}^{N}\left[\int_{\mu_0}^{+\infty} \d\mu\frac{\rho(\mu)}{p_i^2+\mu}-D_i\right]^2+\lambda \int_{\mu_0}^{+\infty} \d\mu~\rho^2(\mu)
\end{equation}
where we use lattice data in momentum space for the gluon propagator computed in a $80^4$ volume, with $\beta=6.0$ (Wilson gauge action) \cite{olisi12}. In eq. (\ref{tikdis1}), $\lambda>0$ is a regularization parameter designed to overcome the ill-posed nature of the inversion. We choose  $\lambda$ by means of the Morozov principle: the optimal value  $\overline\lambda$ is such that the quality of the inversion is equal to the error on the data, i.e.~
$||D^{reconstructed}-D^{data}||=\delta$ where $\delta$ is the total noise on the input data. The IR regulator (threshold) $\mu_0$  will be determined self-consistently by means of the optimal (Morozov) regulator $\overline\lambda$: we take the minimal value for $\overline\lambda(\mu_0)$ that can be reached by varying $\mu_0$. 

The minimization of (\ref{tikdis1})  proceeds through a linear perturbation of $\rho$ and imposing the vanishing of the variation of $\mathcal{J}_\lambda$,
\begin{equation}
   \sum_{i=1}^{N} \underbrace{\left[\int_{\mu_0}^{+\infty}\d\nu\frac{\rho(\nu)}{p_i^2+\nu}-D_i\right]}_{\equiv c_i}\frac{1}{p_i^2+\mu}+\lambda\rho(\mu)=0\,\, (\mu\geq\mu_0)
\end{equation}

The K\"all\'{e}n-Lehmann inverse can be computed explicitly
\begin{equation}
  \rho_{\lambda}(\mu)=-\frac{1}{\lambda}\sum_{i=1}^{N}\frac{c_i}{p_i^2+\mu}\theta(\mu-\mu_0)\,,
\label{specnum}
\end{equation}
where  $\theta(\cdot)$ is the Heaviside step function.  We get a linear system for the coefficients $c_i$: 
\begin{equation}\label{tikdis5}
 \lambda^{-1} \mathcal{M}c+c=-D
\end{equation} 
with
\begin{equation}
  \mathcal{M}_{ij}=\int_{\mu_0}^{+\infty}\d\nu\frac{1}{p_i^2+\nu}\frac{1}{p_j^2+\nu}=\frac{\ln\frac{p_j^2+\mu_0}{p_i^2+\mu_0}}{p_j^2-p_i^2}\,.
\end{equation}

\begin{figure}[t] 
   \centering
   \subfigure[Spectral density.]{ \includegraphics[scale=0.22]{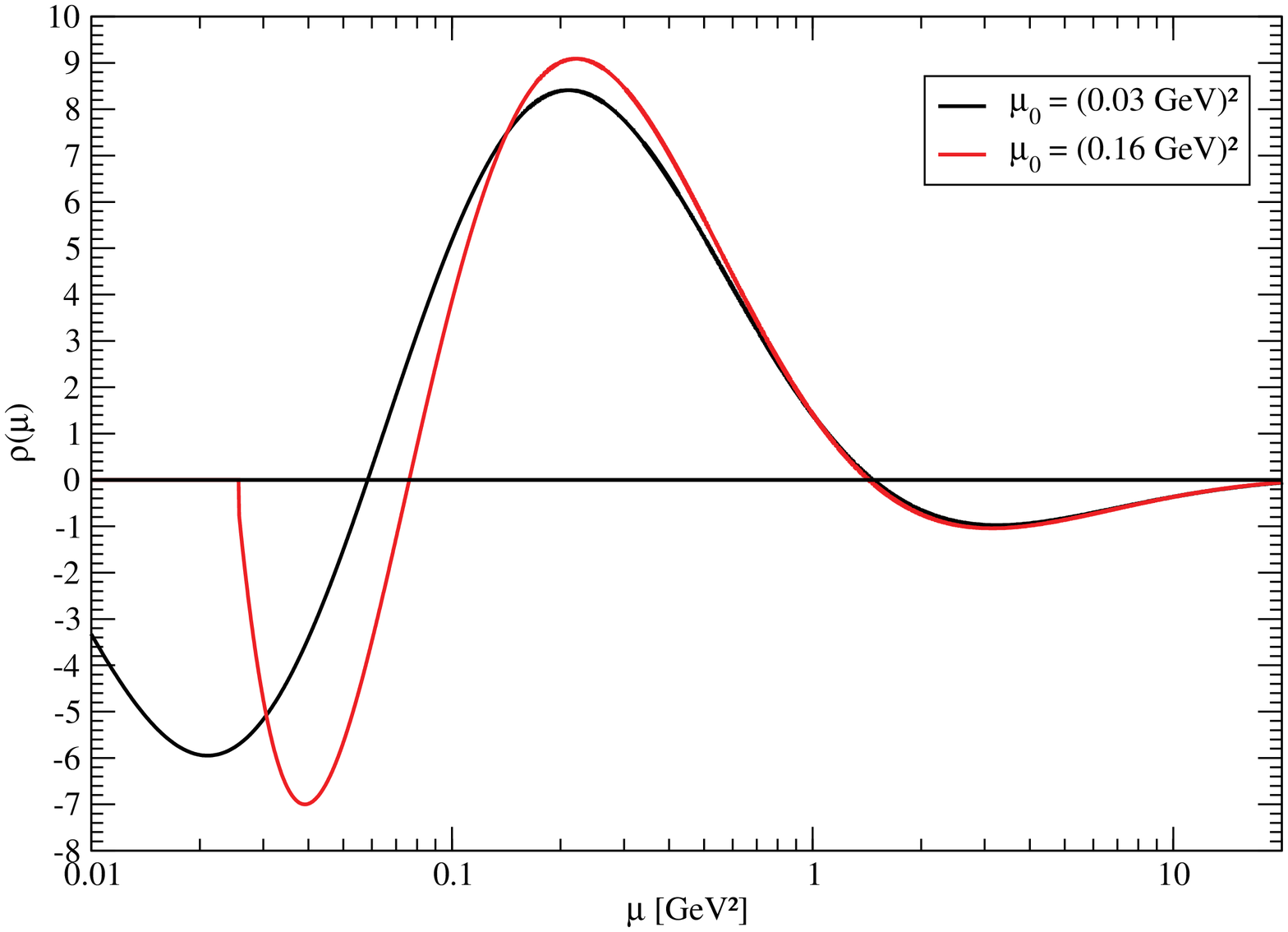} } \quad
   \subfigure[Reconstructed propagator.]{ \includegraphics[scale=0.22]{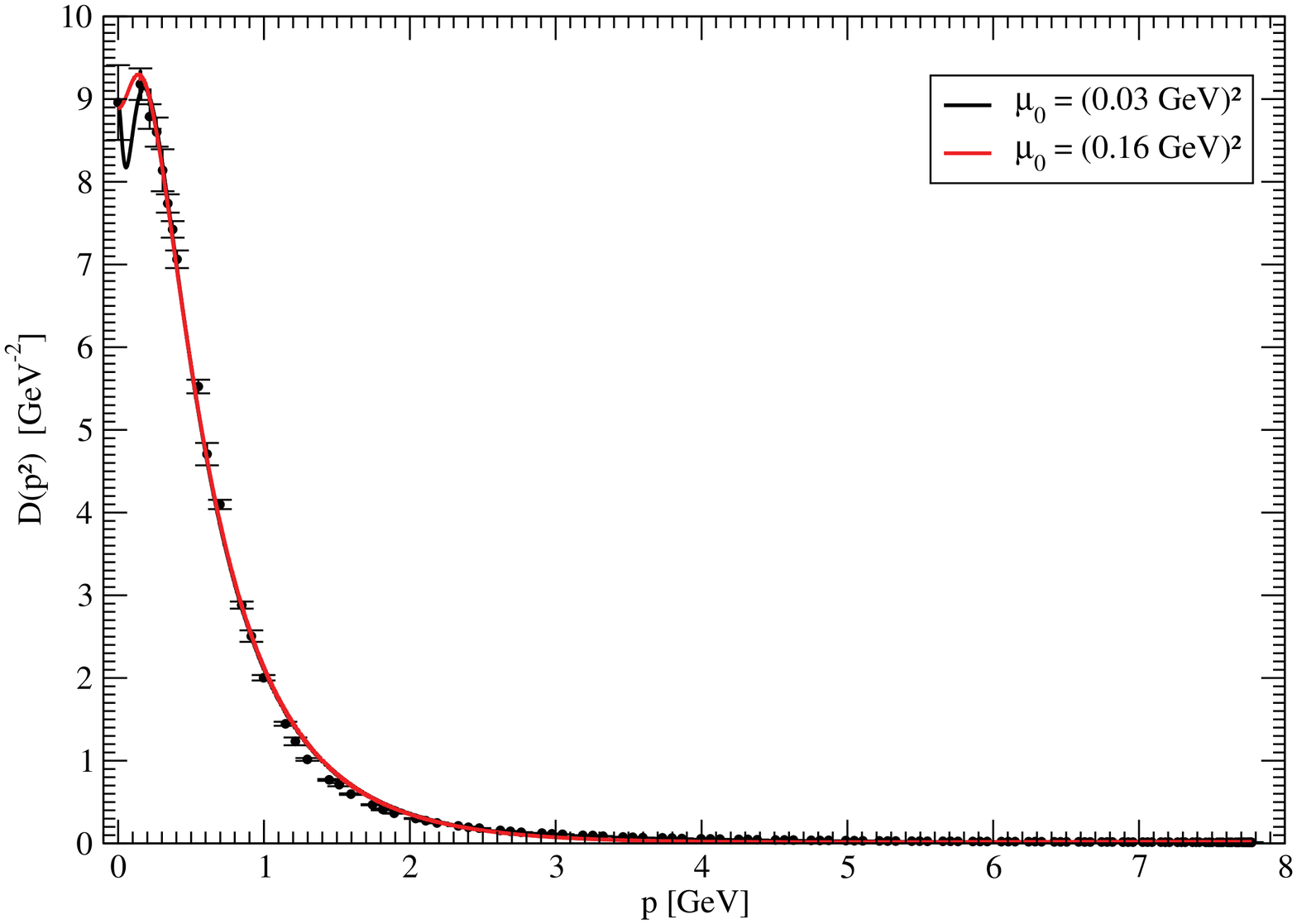} }
  \caption{Results for the gluon spectral function and the reconstructed propagator vs.~the input data. We refer to the main text and \cite{letter} for additional details.}
   \label{spec-plot}
\end{figure}

In Figure \ref{spec-plot} we plot the spectral density and, as a check, the reconstructed propagator which can be easily computed  combining eqs. (\ref{specnum}) and (\ref{specdens}). In this case, we have two minima for  $\overline\lambda(\mu_0)$, at $\mu_0\approx0.03~\textrm{GeV}^2$ and $\mu_0\approx0.16~\textrm{GeV}^2$. We display the results for both values. We conclude that the gluon spectral density is indeed a nonpositive quantity. This is not surprising, since the gluons are not part of the physical spectrum \cite{cornwall}. In the near future, we plan to apply this method to glueballs\footnote{See \cite{latt2012} for a preliminary study of glueball spectral densities.} and other physical degrees of freedom.

\section{Gluon mass at finite temperature}

In this section we briefly describe a recent investigation by some of us 
\cite{gluonmass}, where we address the interpretation of the Landau gauge gluon propagator at finite temperature as a massive-type bosonic propagator. For such a goal, we consider a Yukawa-type propagator
\begin{equation}
 D(p) = \frac{Z }{ p^2 + m^2} \ ,
\label{yukawa}
\end{equation}
where $m$ is the gluon mass and $Z^{\frac{1}{2}}$ the overlap between 
the gluon state and the quasi-particle massive state.

At finite temperature, the Landau gauge gluon propagator is
splitted into two components
\begin{equation}
D^{ab}_{\mu\nu}(\hat{q})=\delta^{ab}\left(P^{T}_{\mu\nu} D_{T}(q_4,\vec{q})+P^{L}_{\mu\nu} D_{L}(q_4,\vec{q}) \right) 
\label{tens-struct}
\end{equation}
where $D_T$ and $D_L$ are the transverse and longitudinal propagators respectively.

\begin{figure}[t] 
   \centering
   \subfigure[Transverse propagator.]{ \includegraphics[scale=0.26]{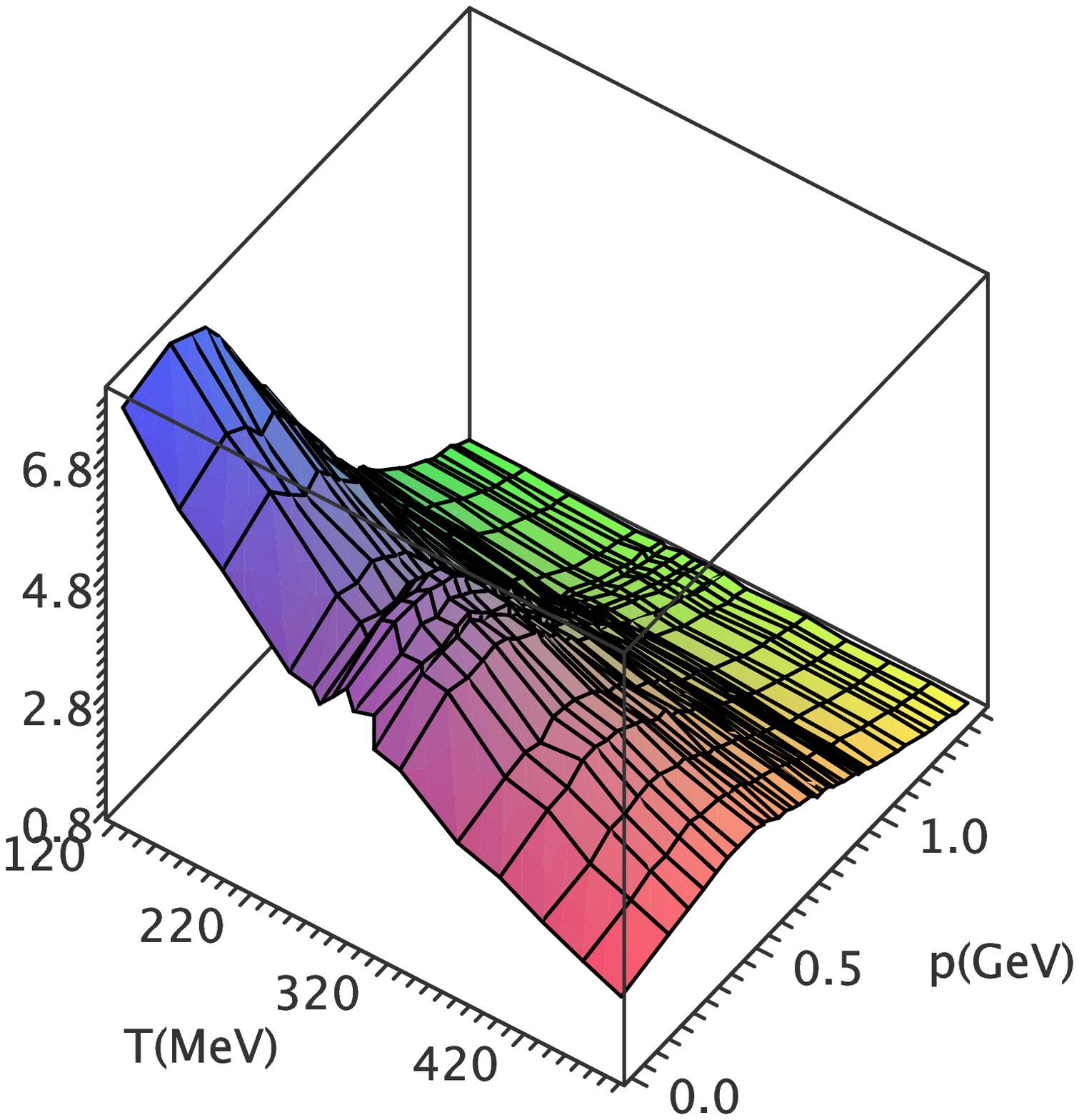} } \quad
   \subfigure[Longitudinal propagator.]{ \includegraphics[scale=0.26]{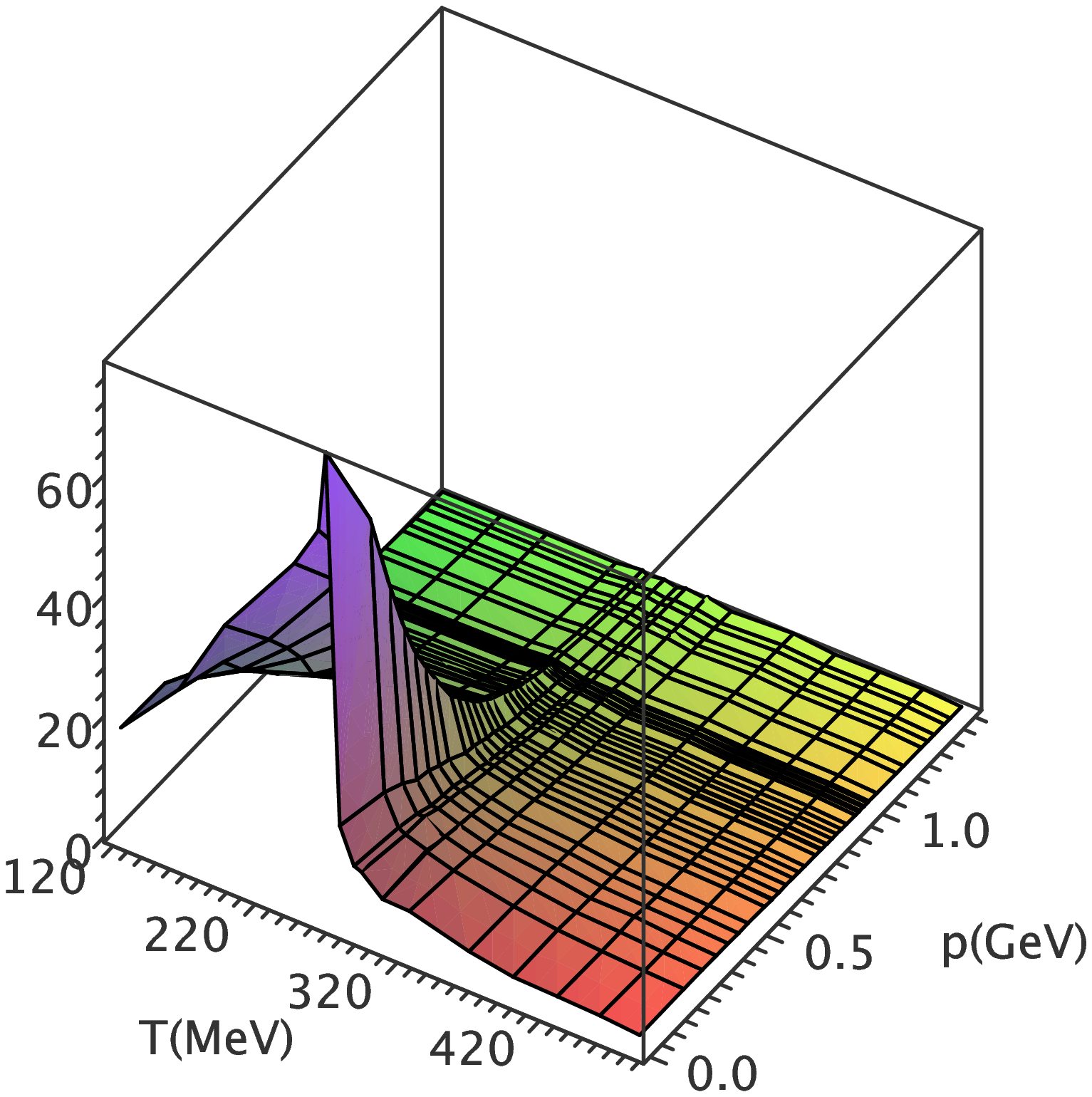} }
  \caption{Components of the gluon propagator as a function of momentum and temperature.}
   \label{tempplot}
\end{figure}

The lattice setup for the simulations at finite temperature considered
here is described in Table \ref{tempsetup}. The surface plots of the two form factors can be seen in Figure \ref{tempplot}. For further details see \cite{gluonmass}.

\begin{table}[t]
\begin{center}
\begin{tabular}{c@{\hspace{0.5cm}}c@{\hspace{0.3cm}}c@{\hspace{0.3cm}}r@{\hspace{0.4cm}}l@{\hspace{0.3cm}}l}
\hline
Temp.            & $\beta$ & $L_s$ &  \multicolumn{1}{c}{$L_t$} & \multicolumn{1}{c}{$a$} & \multicolumn{1}{c}{$1/a$} \\
 (MeV)           &              &            &            & \multicolumn{1}{c}{(fm)} & \multicolumn{1}{c}{(GeV)} \\
\hline
121 &   6.0000 & 64 & 16 & 	0.1016 &  	1.9426 \\
162 &   6.0000 & 64 & 12 & 	0.1016 & 	1.9426 \\
194 &   6.0000 & 64 & 10 & 	0.1016 & 	1.9426 \\
243 &   6.0000 & 64 &   8  & 	0.1016 & 	1.9426 \\
260 &   6.0347 & 68 &   8  & 	0.09502 & 	2.0767 \\
265 &   5.8876 & 52 &   6  & 	0.1243 & 	1.5881 \\
275 &   6.0684 & 72 &   8  & 	0.08974 & 	2.1989 \\
285 &   5.9266 & 56 &   6 & 	0.1154 & 	1.7103 \\
290 &   6.1009 & 76 &   8 & 	0.08502 & 	2.3211 \\
305 &   6.1326 & 80 &   8 & 	0.08077 & 	2.4432 \\
324 &   6.0000 & 64 &   6 & 	0.1016	 &      1.9426 \\
366 &   6.0684 & 72 &   6 & 	0.08974	 &      2.1989 \\
397 &   5.8876 & 52 &   4 & 	0.1243	 &      1.5881 \\
428 &   5.9266 & 56 &   4 & 	0.1154	 &      1.7103 \\
458 &   5.9640 & 60 &   4 & 	0.1077	 &      1.8324 \\
486 &   6.0000 & 64 & 	4 & 	0.1016	 &      1.9426 \\
\hline
\end{tabular}
\end{center}
\caption{Lattice setup used for the computation of the gluon propagator 
at finite temperature. Simulations used the Wilson gauge action; $\beta$ 
was adjusted to have a constant physical volume, $L_s \, a \simeq 6.5$ fm. 
For the generation of gauge configurations and Landau gauge fixing, we used 
Chroma \cite{chroma} and PFFT \cite{pfft} libraries.}
\label{tempsetup}
\end{table}

\begin{figure}[h]
\centering
\includegraphics[scale=0.28]{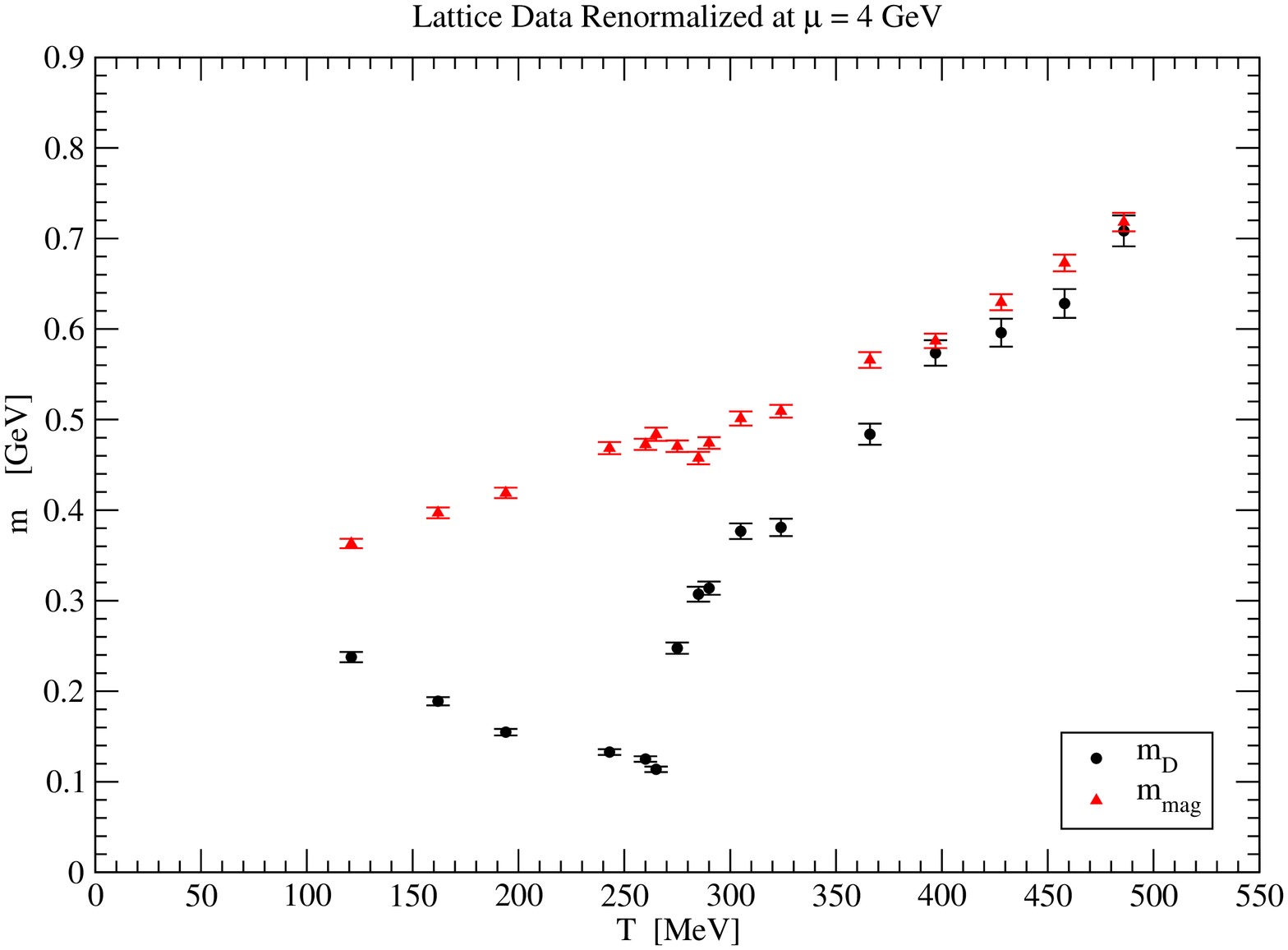}
\caption{Electric and magnetic mass defined from zero momentum propagators. }
   \label{massD0}
\end{figure}

A simple definition for a mass scale associated with the gluon propagator can be given by 
\begin{equation}
 m = 1 / \sqrt{ D(p^2=0; T) } \ .
\label{zeromass}
\end{equation}
Our results for such a mass scale are shown in Figure \ref{massD0}. 

\begin{figure}[h]
\vspace*{0.3cm}
\begin{center}
\includegraphics[width=0.5\textwidth]{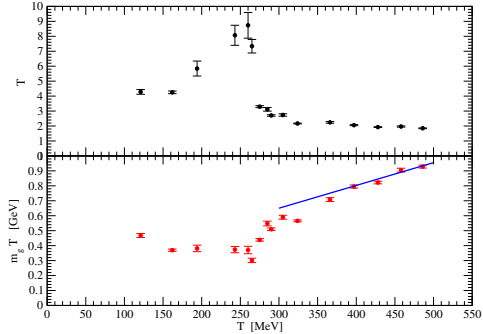}
\caption{$Z(T)$ and $m_g(T)$ from fitting the longitudinal
gluon propagator to a Yukawa form.
The curve in the lower plot is
the fit of $m_g$ to the functional form predicted by perturbation theory.}
\label{yukawafigure}
\end{center}
\end{figure}

A more realistic value for the gluon mass can be obtained by a fit to the 
lattice data in the infrared region using the anzatz described in eq. 
(\ref{yukawa}). It turns out that the transverse form factor is not 
described by a Yukawa-type propagator. Therefore one concludes that $D_T$ 
does not behave as quasi-particle massive boson for $T < 500$ MeV.
In what concerns the longitudinal form factor, we report the values of 
$Z(T)$ and $m_g(T)$ in Figure \ref{yukawafigure}.

We observe that both $m_g(T)$ and $Z(T)$ are sensitive to the confinement-deconfinement phase transition; the data suggests that the phase transition is of first order. Below $T_c$, the gluon mass is a decreasing function of T, whereas it increases for $T>T_c$. Furthermore, the gluon mass follows the expected perturbative behaviour for $T>400$MeV.

\vspace*{-0.1cm}
\section*{Acknowledgments}
Work supported by FCT via projects PTDC/FIS/100968/2008,
CERN/ FP/123612/2011, and CERN/FP/123620/2011
developed under the initiative QREN financed by the UE/FEDER through
the Programme COMPETE - Programa Operacional Factores de Competitividade.
P.~J.~Silva supported by FCT grant SFRH/BPD/40998/2007.
D.~Dudal acknowledges financial support from the Research-Foundation
Flanders (FWO Vlaanderen) via the Odysseus grant of F.~Verstraete. 
Nuno Cardoso supported by NSF award PHY-1212270.


\vspace*{-0.1cm}
\begin{thebibliography}{99}

\bibitem{mem}
M.~Asakawa, T.~Hatsuda and Y.~Nakahara, Prog.\ Part.\ Nucl.\ Phys.\  {\bf 46} (2001) 459.

\bibitem{letter}
D.~Dudal, O.~Oliveira and P.~J.~Silva, Phys.Rev. D {\bf 89} (2014) 014010, arXiv:1310.4069 [hep-lat].

\bibitem{olisi12}
O.~Oliveira and P.~J.~Silva,  Phys.\ Rev.\ D {\bf 86} (2012) 114513, arXiv:1207.3029 [hep-lat].

\bibitem{cornwall} J.~M.~Cornwall, Mod.\ Phys.\ Lett. A {\bf 28} (2013) 1330035, arXiv:1310.7897 [hep-ph].

\bibitem{latt2012} O.~Oliveira, D.~Dudal, P.~J.~Silva, PoS(Lattice 2012)214, arXiv:1210.7794 [hep-lat].

\bibitem{gluonmass}
P.~J.~Silva, O.~Oliveira, P.~Bicudo and N.~Cardoso, Phys.Rev. D {\bf 89} (2014) 074503, arXiv:1310.5629 [hep-lat].

\bibitem{chroma} R.~G.~Edwards and B.~Jo\'o,
Nucl.Phys.Proc.Suppl. {\bf 140} (2005) 832, arXiv:hep-lat/0409003.

\bibitem{pfft} M.~Pippig, SIAM J. Sci. Comput. {\bf 35}, C213 (2013).

\end{thebibliography}
\end{document}